# Epitaxial graphene integrated with a monolayer magnet

Ivan S. Sokolov,[a] Dmitry V. Averyanov,[a] Oleg E. Parfenov,[a] Alexey N. Mihalyuk,[b,c] Alexander N. Taldenkov,[a] Oleg A. Kondratev,[a] Ilya A. Eliseyev,[d] Sergey P. Lebedev,[d] Alexander A. Lebedev,[d] Andrey M. Tokmachev,[a] Vyacheslav G. Storchak[*,a]

[a] National Research Center "Kurchatov Institute", Kurchatov Sq. 1, 123182 Moscow, Russia

[b] Institute of High Technologies and Advanced Materials, Far Eastern Federal University, 690950 Vladivostok, Russia

[c] Institute of Automation and Control Processes FEB RAS, 690041 Vladivostok, Russia

[d] Ioffe Institute, 194021 St. Petersburg, Russia

## Abstract

Imprinting magnetism into graphene makes an important step to its applications in spintronics. An actively explored approach is proximity coupling of graphene to a 2D magnet. In these endeavors, the use of epitaxial graphene may bring significant advantages due to its superiority over the exfoliated counterpart and natural integration with the substrate but the problem of attaining magnetism persists. Here, we report synthesis and analysis of a heterostructure coupling epitaxial graphene with a regular lattice of magnetic atoms formed by Eu intercalation. The magnetization measurements reveal easy-plane 2D magnetism in the material, with the transition temperature controlled by low magnetic fields. The emerging negative magnetoresistance and anomalous Hall effect point at spin polarization of the carriers in graphene. In the paramagnetic phase, the magnetoresistance in graphene exhibits critical exponential behavior of the induced magnetic state. The intercalation does not compromise the parental electronic structure – quantum oscillations in the resistivity manifest low-mass carriers in graphene. The results are set against those for an isostructural material based on intercalated nonmagnetic Sr. Overall, the study expands the family of 2D magnets and establishes a prospective material for graphene-based spintronics.

---

[*] Corresponding author. *E-mail address:* vgstorchak9@gmail.com (Vyacheslav Storchak)

# 1. Introduction

The demands for ultra-compact spintronics drive research into the 2D limit of magnetism. Recently, 2D magnets have emerged as a rich source of unconventional phases and exceptional phenomena [1-3]. They are particularly promising for applications in spintronics, including optospintronics [4,5]. The 2D structure of the magnets makes them natural components in various functional heterostructures. The practical applications take advantage of the high amenability of the magnetic order in 2D systems to external stimuli [6] such as magnetic fields [7], strain [8] or gating [9]. The coupling of magnetic order to electron transport gives rise to extraordinary magnetotransport effects [10,11]. In addition, the studies of 2D magnets deepen our understanding of fundamentals of magnetism, provide valuable insights into spin dynamics [12], magnetic anisotropy [13], and magnon phenomena [14]. Remarkably, the attained results are based on a very small pool of available 2D magnets. Further developments hinge upon expansion of this pool, making the materials issue central to this area of research. Most of 2D magnets derive their magnetic properties from open $d$-shells of transition metal atoms such as Cr [7] and Fe [9] or open $f$-shells of rare earth metals such as Eu and Gd [15,16]. However, the traditional 2D materials, not containing those elements, are often left aside in the research on 2D magnetism, despite the wealth of advantageous properties that can be of use in spintronic devices. In particular, graphene (Gr), the poster child of 2D materials science, can bring a lot to the table.

A long spin lifetime and weak hyperfine interactions make graphene a prospective material for spintronics [17,18] – various spin logic elements based on graphene are devised [19]. The chemical stability of graphene allows for its use as a mediator of spin dynamics in heterostructures [20,21]. However, the applications of graphene in spintronics are limited by its nonmagnetic character. Therefore, significant efforts have been invested into making graphene magnetic [22]. One approach is to induce imbalance between the two sublattices of graphene which can be done by adatoms [23] or edge effects in nanostructures [24]. Also, magnetic states can emerge in moiré graphene systems [25]. The most straightforward approach is to couple graphene with a magnetic material. Such designer heterostructures have been fabricated for various magnetic insulators, mostly oxides [26-29], and shown to induce spin polarization in graphene. The advent of 2D magnets initiated research into magnetic graphene heterostructures where all the components are 2D materials. At present, the majority of prominent 2D magnets, both ferromagnetic (FM) and antiferromagnetic (AFM), have been integrated with graphene: the list includes $CrI_3$ [30], $CrBr_3$ [31], $Cr_2Ge_2Te_6$ [32-34], $Fe_3GeTe_2$ [35,36], $Fe_3GaTe_2$ [37], CrSBr [38,39], $GdSi_2$ [40], $MnPSe_3$ [41], etc. The heterostructures are not limited to those of a single-layer graphene [32,42]. Moreover, non-van der Waals monolayers can also be used for magnetic proximity effects in graphene [43,44]. The graphene heterostructures exhibit various spin-related effects such as giant magnetoresistance (MR) [32,35], anomalous Hall effect (AHE) [33,34,39,40], spin Hall effect [31,34], and topological Hall effect [36]; they serve as components of spintronic devices [30,34,37,45]. Despite those achievements, the research is still far from large-scale applications; it encounters significant challenges regarding the scalability of the materials synthesis and their integration with a substrate.

Epitaxial graphene systems, formed on SiC, have key advantages over their exfoliated counterpart resulting from the uniform growth processes and compatibility with the semiconductor fabrication technologies [46,47]. The properties of graphene can be altered by metal intercalation [46,48]. At the same time, the graphene layer protects the metal layer stabilized at the interface with the SiC substrate. Intercalation of an inherently magnetic metal may provide the coveted magnetic response in graphene.

A crucial issue is the structural quality of the resulting heterostructure – ideally, graphene should be coupled with a *crystalline* monolayer magnet. Regretfully, magnetic metal intercalation into the confined space between graphene and SiC typically results in metal clusters and islands, irrespective of whether transition metals [49] or rare earths [50] are used. However, the lack of systematic overarching studies including optimization of the synthesis conditions suggests that the problem may still have a solution.

To reach a positive outcome, the choice of the metal for intercalation is highly important: it should preferably be a metal with a low affinity to covalent bonding and known to form a stable stoichiometric graphite intercalation compound. Europium, an active magnetic metal forming the $EuC_6$ compound in the bulk [51] and few layers [52], satisfies these conditions. Indeed, various Gr/Eu systems have been produced [53-59]. Eu can be adsorbed on top of graphene to form a monolayer with the $EuC_6$ stoichiometry but the system is unstable – its magnetic properties depend strongly on the capping layer [55]. Intercalation of Eu into the space between graphene and semiconducting Si produces a surface reconstruction of Eu on the Si substrate, incommensurate with graphene [58]. The main problem of Gr/Eu/M (M = metal) systems [53,56,57,59] is that the shunting metal substrate prevents studies of electron transport and hinders applications of such materials. Eu may aggregate into clusters and islands [53]. Moreover, the metal substrate affects the magnetic properties of the heterostructure, especially in the case of FM metals Co [56,57] and Ni [56,59]. Most importantly, the problem with the structural quality applies to Eu intercalation under epitaxial graphene on SiC: Eu nanoclusters with the AFM properties of bulk Eu are formed [54]. Therefore, the prospects of Gr/Eu magnets in nanoelectronics are still an open question.

Here, we meet the challenge by integrating epitaxial graphene with a monolayer lattice of Eu on SiC, equivalent to a $EuC_6$ monolayer. The heterostructure is probed by magnetization measurements to determine that it is a 2D magnet. Electron transport studies signify spin polarization of low-mass carriers in graphene. In the paramagnetic phase, the carrier transport exhibits a critical behavior. The results are compared to those for an equivalent heterostructure formed by nonmagnetic Sr.

## 2. Experimental section

### 2.1. Synthesis

Epitaxial graphene was produced by the SiC surface sublimation method [60]. High-purity semi-insulating 4H-SiC(0001) wafers (TanKeBlue Semiconductor Co., Ltd.) were employed for the synthesis; the miscut angle did not exceed 0.25°. The substrate surface was prepared by chemical-mechanical polishing. Then, epitaxial graphene was synthesized in a high-frequency induction heating system furnished with graphite crucible: the SiC wafer was heated in a chamber filled with 6N Ar ($750 \pm 20$ Torr) up to $1730 \pm 20$ °C at a rate 100 °C/min and then annealed for 5 min. The substrate was transferred into a Riber Compact molecular beam epitaxy (MBE) system maintaining the base pressure in the growth chamber below $10^{-10}$ Torr. It was then annealed for 10 min at 400 °C. Gr/Eu/SiC was synthesized by deposition of Eu at 250 °C for 90 s. The metal was supplied from a Knudsen cell effusion source heated to 430 °C; according to a Bayard-Alpert ionization gauge, it corresponded to a Eu partial pressure of $2.5 \cdot 10^{-8}$ Torr. Similarly, Gr/Sr/SiC was synthesized by heating the corresponding effusion source to 240 °C ($P(Sr) = 1 \cdot 10^{-8}$ Torr) and deposition of Sr at 150 °C for 80 s. To avoid degradation by air, each sample was capped with a 20-nm layer of amorphous $SiO_x$ deposited at room temperature.

### 2.2. Characterization

The surface structures of the samples were continuously monitored in the MBE chamber using reflection high-energy electron diffraction (RHEED) spectra recorded with a RHEED diffractometer equipped with the kSA 400 analytical system. Outside the MBE system, grazing incidence X-ray diffraction (GIXRD) techniques were employed to characterize the atomic structures of the samples: GIXRD scans were recorded in a Rigaku SmartLab 9 kW diffractometer operating at a wavelength of 1.54056 Å (Cu $K_{\alpha 1}$). The magnetic properties of the materials were measured with an MPMS XL-7 magnetometer employing the reciprocating sample option; samples with a lateral size of 5 mm were aligned with respect to the applied magnetic field with an accuracy of better than 2°. The diamagnetic contribution from the substrate was subtracted as in the case of the 2D magnet $EuC_6$ [55]. The lateral electron transport measurements were carried out using the LakeShore 9709A system. Four-contact measurements, adhering to the ASTM standard F76, employed 5 mm × 5 mm square samples. The ohmic electrical contacts were fabricated by deposition of an Ag-Sn-Ga alloy onto each terminal; their quality was attested by I-V characteristic curves.

### 2.3. Computational techniques

The DFT calculations were carried out with the Vienna *ab initio* simulation package (VASP) [61] employing the projector-augmented wave approach [62] and the generalized gradient approximation (GGA) of Perdew, Burke, and Ernzerhof (PBE) [63] as the exchange-correlation functional. The substrate was simulated by a slab consisting of six bilayers of 4H-SiC(0001) with a PBE-optimized bulk lattice constant ($a = 3.068$ Å). The positions of the atoms in the two bottom SiC bilayers were fixed and the rest was relaxed until the atomic forces were below 0.01 eV/Å. The dangling bonds at the bottom of the slab were passivated by H atoms to emulate the bulk behavior. The vacuum space between the slabs was set to 1.5 nm. The Gr/M/SiC systems were simulated using a 3 × 3 SiC supercell that accommodates four $\sqrt{3} \times \sqrt{3}$ unit cells of graphene layers. In this case, the lattice mismatch was about 7.7 %, known to provide a reasonable approximation to the electronic structure of the system [64]. The plane-wave cut-off energy was set to 400 eV; the Brillouin zone integration was carried out with a Gaussian smearing of 10 meV. A 6 × 6 × 1 Γ-centered $k$-point mesh was used to sample the supercell's Brillouin zone. The band structures of Gr/M/SiC were calculated using a spin-polarized noncollinear regime. A Hubbard $U$ correction was applied to the $f$-shells of Eu according to the Dudarev's method [65]. Its value, 5.0 eV, was chosen to agree with the results on the position of the Eu $f$-band produced with the Heyd-Scuseria-Ernzerhof (HSE06) screened hybrid functional [66]. The scalar relativistic effect and the spin-orbit coupling were taken into account.

## 3. Results and discussion

### 3.1. Synthesis and structural characterization

To reach the goals of the study, it is necessary to determine the systems to be synthesized and a suitable synthetic technique. The surface of SiC with an epitaxial graphene overlayer is rather complex. Below graphene, there is the so-called buffer layer, a graphene-like carbon layer covalently bound to the SiC substrate. Therefore, one can imagine 3 major types of Eu layer locations with respect to the Gr/SiC system: (i) Eu adsorbed on top of the graphene layer; (ii) Eu intercalated between the graphene and buffer layers; (iii) Eu intercalated underneath the buffer layer. Our aim is to produce a heterostructure of type (ii) because of the following two prerequisites: Eu is in direct contact with graphene and the latter protects Eu from the environment. Computational analysis of the 3 types of Eu structures may shed light on their relative stability but is challenging because the lattice of graphene is incommensurate to bulk SiC. Nevertheless, we carried out density functional theory (DFT) calculations of the band structures and energies for toy models, with the $EuC_6$ stoichiometry, representing the 3 types of structures (see Experimental section). The calculations suggest that the target material has the lowest energy (Fig. S1). In the following, it is denoted as Gr/Eu/SiC.

To distinguish the effects related to the open 4$f$-shell of Eu, it is desirable to set the results on Gr/Eu/SiC against those of a reference nonmagnetic system. In this regard, it is possible to use the relationship between Eu and Sr. The $Eu^{2+}$ and $Sr^{2+}$ ions have close radii; as a rule, Eu compounds are isostructural to their Sr analogues. The main difference is that Sr is nonmagnetic. Magnetic Eu compounds are often compared to their Sr counterparts, including the studies of Eu-based 2D magnets [67]. Similar to $EuC_6$, the isostructural $SrC_6$ material is produced by Sr intercalation into graphite [68]. Therefore, we complement the study of Gr/Eu/SiC with a study of Gr/Sr/SiC of the $SrC_6$ stoichiometry. Our DFT calculations suggest that the structure with Sr atoms intercalated between the graphene and buffer layers has the lowest energy, similar to the case of Eu-based materials (Fig. S1). As far as we are aware, synthesis of 2D analogues of $SrC_6$ is yet to be reported.

2D magnetic heterostructures can be produced by different techniques ranging from manual stacking to chemical vapor deposition to MBE [69]. Our choice here is MBE, a bottom-up approach offering precise control over the synthesis on the monolayer level. Moreover, MBE has been instrumental in synthesis of a number of Eu-based 2D magnetic systems [15,16,67], including those based on graphene [40,52,55,58]. To synthesize

Gr/Eu/SiC and Gr/Sr/SiC, we employed 4H-SiC wafers and fabricated the layer of epitaxial graphene by the surface sublimation method (see Experimental section). Then, the heterostructures were synthesized by deposition of Eu or Sr, respectively. The substrate was held at an elevated but rather modest temperature. The temperature regime is extremely important: at higher temperature, Eu is known to intercalate under the buffer layer [70]. The amounts of deposited metals were carefully tuned. Although the graphene layer protects the intercalated metals, their oxidation by air may still be a problem because of the high chemical reactivity of both Eu and Sr. Therefore, the heterostructures were capped with a layer of amorphous $SiO_x$, a nonmagnetic insulator.

To characterize the atomic structure of the materials we employ a combination of techniques. We studied materials before the synthesis, *i.e.* the Gr/SiC system, and the changes caused by Eu intercalation. First, the sample was monitored *in situ* employing RHEED. This technique is sensitive to the surface, which makes it particularly useful for studies of 2D materials. The complex atomic structure of the Gr/SiC system results in a rather complex RHEED pattern (Fig. 1a), which demonstrates signals from both graphene and the SiC substrate. Intercalation of Eu modifies the RHEED image – new reflexes appear (Fig. 1c). These reflexes correspond to a $(\sqrt{3} \times \sqrt{3})$ $R30°$ superstructure, well known for metal atoms intercalated under graphene. This observation agrees with the formation of a triangular monolayer lattice of Eu coupled to graphene. The resulting material corresponds to the structure where each third $C_6$ ring of graphene accommodates a Eu atom. In principle, Eu in contact with graphene may form an alternative, $2 \times 2$, phase reported for incomplete intercalation of Eu into the gap between graphene and Ir(111) [71]. However, we do not observe any signs of such a phase.

The rest of the structural characterization was carried out *ex situ*. To confirm the stability of the structure with respect to the $SiO_x$ deposition and exposure to air, GIXRD, a technique instrumental for structural studies of 2D materials, was employed. Fig. 1b shows the GIXRD scan for the substrate, revealing peaks from SiC and epitaxial graphene. In Gr/Eu/SiC, new peaks emerge (Fig. 1d); they are identified as stemming from the $(\sqrt{3} \times \sqrt{3})$ $R30°$ lattice observed in the RHEED spectra. No side phases are detected. The results are confirmed by the GIXRD scan along a different direction (Fig. S2). The stability of the structure against the $SiO_x$ capping layer may serve as an indication that the Eu atoms are intercalated rather than adsorbed on the graphene surface. Raman spectra of the material before and after intercalation are presented in Fig. S3. Based on these data, we do not observe any significant increase in the defects concentration. However, the intensity of the 2D line after intercalation has decreased significantly, which points to an increase in charge carrier concentration [72].

## 3.2. Magnetic properties

First, it is necessary to establish the magnetic properties of the heterostructure. The magnetic signals of monolayer magnets are weak; special techniques are often used for their detection. However, the MBE synthesis employing epitaxial graphene on SiC provides large-area (cm-size) Gr/Eu/SiC samples; therefore, SQUID magnetization measurements are sufficiently sensitive to carry out the study. Fig. 2a demonstrates temperature dependence of the FM moment in different in-plane magnetic fields. The emergence of an FM moment in Gr/Eu/SiC is in sharp contrast to the magnetic properties of samples with Eu adsorbed on graphene: according to Ref. [55], Eu/Gr capped with $SiO_x$ does not demonstrate any detectable FM signals. A striking feature of the $M_{FM}(T)$ curves in Fig. 2a is a strong dependence of the effective Curie temperature on weak magnetic fields, which is a hallmark of 2D magnetism. It arises because the magnetic fields control the pseudogap in the spin-wave excitation spectrum [73]. Such a dependence has been observed in various 2D FM materials, from $Cr_2Ge_2Te_6$ [74] to a selection of rare-earth 2D magnets [15,16], including those based on graphene [52,55,58]. In contrast to Gr/Eu/SiC, the reference material Gr/Sr/SiC does not show any FM signals in the $M_{FM}(T)$ dependence, as illustrated by Fig. S4.

The magnetic field dependence of the magnetic moment in Gr/Eu/SiC follows an *S*-type curve (Fig. 2b). In contrast to other systems combining Eu and monolayer graphene [54,55,58,71], Gr/Eu/SiC demonstrates a significant hysteresis loop (Fig. 2b). The emergence of the magnetic moment is associated with the lattice of Eu atoms, as follows from the comparison of the data for Gr/Eu/SiC and Gr/Sr/SiC (Fig. 2b). As expected, the $M_{FM}(H)$ dependence in Gr/Eu/SiC vanishes as temperature increases (Fig. S5). The area-normalized saturation magnetic moment in Gr/Eu/SiC is close to that in monolayer $CrI_3$ [75], probably the best studied 2D magnet. However, the saturation magnetic moment is far from the value expected for fully spin-polarized Eu atoms carrying a magnetic moment of 7 $\mu_B$ per atom. This situation is not unusual – various 2D magnets based on Eu and Gd with the $4f^7$ electron configuration exhibit significantly reduced magnetic moments [15,16,55,67,76]. This fact is explained by a competition between FM and AFM orders in the materials. The same conclusion is drawn for Eu adsorbed on the surface of graphene [77]. A recent study of the 2D magnet $EuGe_2$, a material with the triangular lattice of Eu similar to Gr/Eu/SiC, images a magnetic phase separated state [78]. The complexity of the magnetic state hinders its computational analysis – DFT calculations of 2D magnets with the triangular lattice of Eu, EuSi$_2$ [79] and EuGe$_2$ [80] do not capture the salient features of their magnetic structures. Besides the reduced magnetic moment, one of those features is easy-plane anisotropy of the magnetism. The same type of magnetic anisotropy is observed in Gr/Eu/SiC; it follows from both temperature (Fig. 2c) and field (Fig. S6) dependences of the magnetic moment in magnetic fields of different directions, in-plane and out-of-plane. The emergence of the FM state in Gr/Eu/SiC is further supported by measurements of the remnant moment (Fig. 2d) and the FC/ZFC bifurcation (Fig. S7).

### 3.3. Electron transport properties

The main purpose of constructing the lattice of Eu atoms is to induce spin polarization of the carriers in the adjacent graphene layer. The band structure calculation of Gr/Eu/SiC (Fig. S8) demonstrates that the electron states in proximitized graphene are spin polarized, in contrast to those in Gr/Sr/SiC. The electron states of graphene can be probed by electron transport measurements because the SiC substrate is an insulator. Fig. S9 shows temperature dependence of resistance in the two materials. The curve for Gr/Eu/SiC exhibits a significant increase in the resistance at low temperature. The evolution of the feature in $R(T)$, associated with the magnetic transition, in magnetic fields may provide information on the magnetic state in graphene. Fig. S10 compares the $R(T)$ dependences for Gr/Eu/SiC in zero magnetic field and a magnetic field 9 T parallel to the current. The shift of the feature to a lower temperature in the magnetic field indicates that the predominant magnetic order in Gr/Eu/SiC is AFM. This shift is particularly well seen in the temperature dependence of the derivative of the resistance (Fig. 3a). This plot also helps to determine that $T_c$ is 26.5 K. It should be noted that the same derivative in the case of Gr/Sr/SiC is featureless (Fig. 3a).

The electron transport properties of Gr/Eu/SiC differ significantly from those of pristine graphene. Fig. 3b demonstrates negative MR at low temperatures; to avoid localization effects, the MR is shown for magnetic fields parallel to the current. The negative MR may indicate spin polarization of carriers in graphene; its absolute value decreases and then vanishes as temperature increases. The MR is anisotropic; in particular, there is a difference between the MR for in-plane magnetic fields parallel and orthogonal to the current. Fig. 3c provides the temperature dependence of the MR anisotropy in Gr/Eu/SiC and Gr/Sr/SiC. At high temperatures, the difference between the two materials is marginal. However, around $T_c$ the MR anisotropy dependences for Gr/Eu/SiC and Gr/Sr/SiC diverge. Therefore, the MR anisotropy is sensitive to the magnetic transition. In general, the coupling of graphene to a 2D magnet is expected to result in doping of graphene. Some 2D magnets are known for hole doping in heterostructures with graphene [81,82] but Eu, being an active metal, should be a source of electron doping. Indeed, this conclusion is supported by analysis of the linear contribution to the Hall effect. In the case of Gr/Eu/SiC, the Hall effect dependence on the applied magnetic field also contains a nonliear contribution attributed to the AHE (Fig. 3d). For comparison, the nonlinearity of the $R_{xy}(H)$ dependence in Gr/Sr/SiC is negligible (Fig. 3d). The AHE is yet another evidence of the spin polarization of carriers in graphene. The AHE in Gr/Eu/SiC depends strongly on temperature (Fig. S11), becomes very small at temperatures above $T_c$.

Fig. 3b suggests that the negative MR in Gr/Eu/SiC is significant even in the paramagnetic (PM) region. Actually, the temperature dependence of the MR in the PM region (Fig. 4a) reminds of a critical behavior. It provides a way to determine a critical exponent for magnetic states induced in graphene. The MR can be associated with a local magnetic susceptibility $\chi_{loc}$, probed by electron; more precisely, the MR is proportional to $(\chi_{loc}H)^2$. Indeed, the main contribution to the MR in the PM region can be roughly approximated by a parabola (see Fig. 4b for an illustration). Based on these dependences, we can estimate how $\chi_{loc}$ changes with temperature. Its critical behavior suggests that $\chi_{loc}$ is proportional to $(T + T_c)^{-\gamma}$; the sign takes into account the AFM character of the magnetic state, as demonstrated above. Fig. 4c shows that the determined $\chi_{loc}$ does follow this behavior. The determined critical exponent $\gamma = 1.76(4)$. The data discussed above are produced for the MR in a magnetic field parallel to the current. However, it would be difficult to justify if the behavior of $\chi_{loc}$ depended on the current. Fig. 4d shows that this is not the case: the dependence of $\chi_{loc}$ on temperature is the same for in-plane magnetic fields parallel and perpendicular to the current. The determination of $\gamma$ in Gr/Eu/SiC is important because experimental studies of the critical exponents in 2D magnets are very difficult [13]; we are not aware of any such studies for magnetic graphene.

We showed the effects of intercalated Eu on the electron transport in graphene. However, it is important to establish that the advantageous properties of graphene are not ruined in the heterostructure. After all, this is one of the main rationales behind the use of Eu rather than transition metals to induce magnetism in graphene [56,71]. In this regard, quantum (Shubnikov-de-Haas, SdH) oscillations can be quite informative. Fig. 5a demonstrates SdH oscillations in the resistance of Gr/Eu/SiC. For comparison, SdH oscillations in Gr/Sr/SiC are shown in Fig. S12. The periods of the oscillations in the two systems are very close: the estimated carrier concentrations in Gr/Eu/SiC and Gr/Sr/SiC are $1.06 \cdot 10^{13}$ cm$^{-2}$ and $1.10 \cdot 10^{13}$ cm$^{-2}$, respectively. It justifies the use of Sr for comparison with Eu in the intercalated materials. The carrier concentration in Gr/SiC, extracted from SdH oscillations, is much lower, about $5.4 \cdot 10^{11}$ cm$^{-2}$, signifying a noticeable doping of graphene by Eu and Sr. It agrees with the significant reduction of the intensity of the 2D line in the Raman spectra after intercalation. The amplitudes of the oscillations decrease with temperature (Fig. 5b). The decrease is slightly steeper in the case of Gr/Eu/SiC suggesting a higher effective mass. Based on the Lifshitz-Kosevich formalism, the estimated effective masses of carriers in Gr/Eu/SiC and Gr/Sr/SiC are 0.084(2) $m_e$ and 0.074(2) $m_e$, respectively. In both cases, the masses are rather low, to compare with the system Gr/Gd/Si, where the proximity effects resulted in the mass renormalization to 0.2 $m_e$ [43]. One of the reasons for a slightly higher mass in Gr/Eu/SiC may be renormalization due to magnetic interactions. The values of the masses are in good correspondence to the square-root dependence of the cyclotron mass on the carrier concentration in graphene [83]. It suggests that the electronic structure of graphene is preserved in the intercalated materials.

## 4. Conclusion

Rare earths are instrumental in numerous technologies and devices but their potential in 2D magnetism is yet to be appreciated. Among rare earths, europium is of particular interest for magnetic materials because of its high local magnetic moments and chemical activity. They make Eu a prospective candidate to solve the long-standing problem of inducing magnetism in graphene *via* chemical functionalization. In particular, the chemical activity of Eu gives it advantage over transition metals that tend to form covalent bonds and destroy the electronic structure of graphene. Here, we managed to synthesize a heterostructure combining epitaxial graphene on SiC with the intercalated monolayer lattice of Eu atoms, overcoming the difficulties of previous attempts that resulted in nanoclusters of Eu. The material, Gr/Eu/SiC, turns out to be a magnet with the EuC$_6$ stoichiometry sharing magnetic characteristics, such as easy-plane magnetic anisotropy and reduced magnetic moments, with other Eu-based 2D magnets. The 2D nature of the magnetic states follows from the high sensitivity of the effective $T_c$ to weak magnetic fields. Importantly, the proximity to the Eu lattice induces spin polarization of carriers in graphene, as witnessed by the negative MR and the AHE. The analysis of the SdH oscillations in resistivity suggests that, despite the induced spin polarization and significant doping, graphene largely preserves its identity. The stoichiometric nonmagnetic material Gr/Sr/SiC, designed and synthesized for comparison with Gr/Eu/SiC, may be of interest in its own right. The Gr/Eu/SiC material may be employed as a part in graphene-based van der Waals heterostructures. Hopefully, the magnetized epitaxial graphene will find applications in ultra-compact spintronics.

## CRediT authorship contribution statement

**Ivan S. Sokolov:** Investigation. **Dmitry V. Averyanov:** Investigation. **Oleg E. Parfenov:** Investigation. **Alexey N. Mihalyuk:** Investigation. **Alexander N. Taldenkov:** Investigation. **Oleg A. Kondratev:** Investigation. **Ilya A. Eliseyev:** Investigation. **Sergey P. Lebedev:** Investigation. **Alexander A. Lebedev:** Investigation. **Andrey M. Tokmachev:** Writing – Original Draft. **Vyacheslav G. Storchak:** Investigation, Writing – Original Draft, Supervision.

## Declaration of competing interest

The authors declare that they have no known competing financial interest or personal relationships that could have appeared to influence the work reported in this paper.

## Acknowledgments

This work was supported by NRC "Kurchatov Institute" (structural characterization) and the Russian Science Foundation [grant No. 24-19-00038 (synthesis, studies of magnetism and electron transport)]. The measurements were carried out using equipment of the resource centers of electrophysical and laboratory X-ray techniques at NRC "Kurchatov Institute". The calculations were carried out using equipment of the resource center "Far Eastern Computing Resource" IACP FEB RAS (https://cc.dvo.ru).

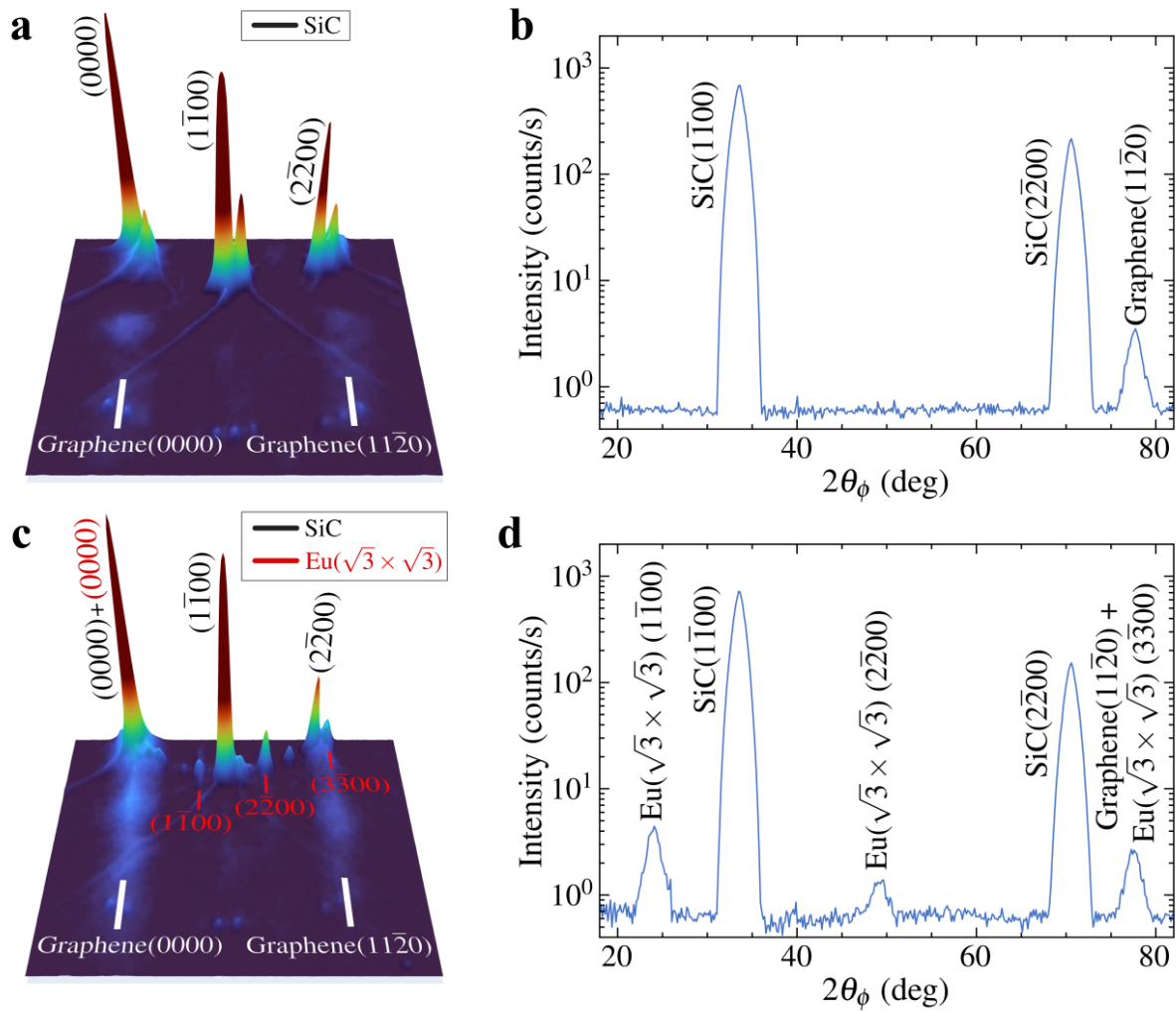

**Fig. 1.** a) 3D RHEED image of epitaxial graphene on SiC; the major reflexes are indicated for graphene (white) and SiC (black). b) Grazing incidence XRD scan of epitaxial graphene on SiC along the direction [1$\bar{1}$00] of the substrate. c) 3D RHEED image of Gr/Eu/SiC; the major reflexes are indicated for graphene (white), SiC (black), and the Eu lattice (red). d) Grazing incidence XRD scan of Gr/Eu/SiC along the direction [1$\bar{1}$00] of the substrate.

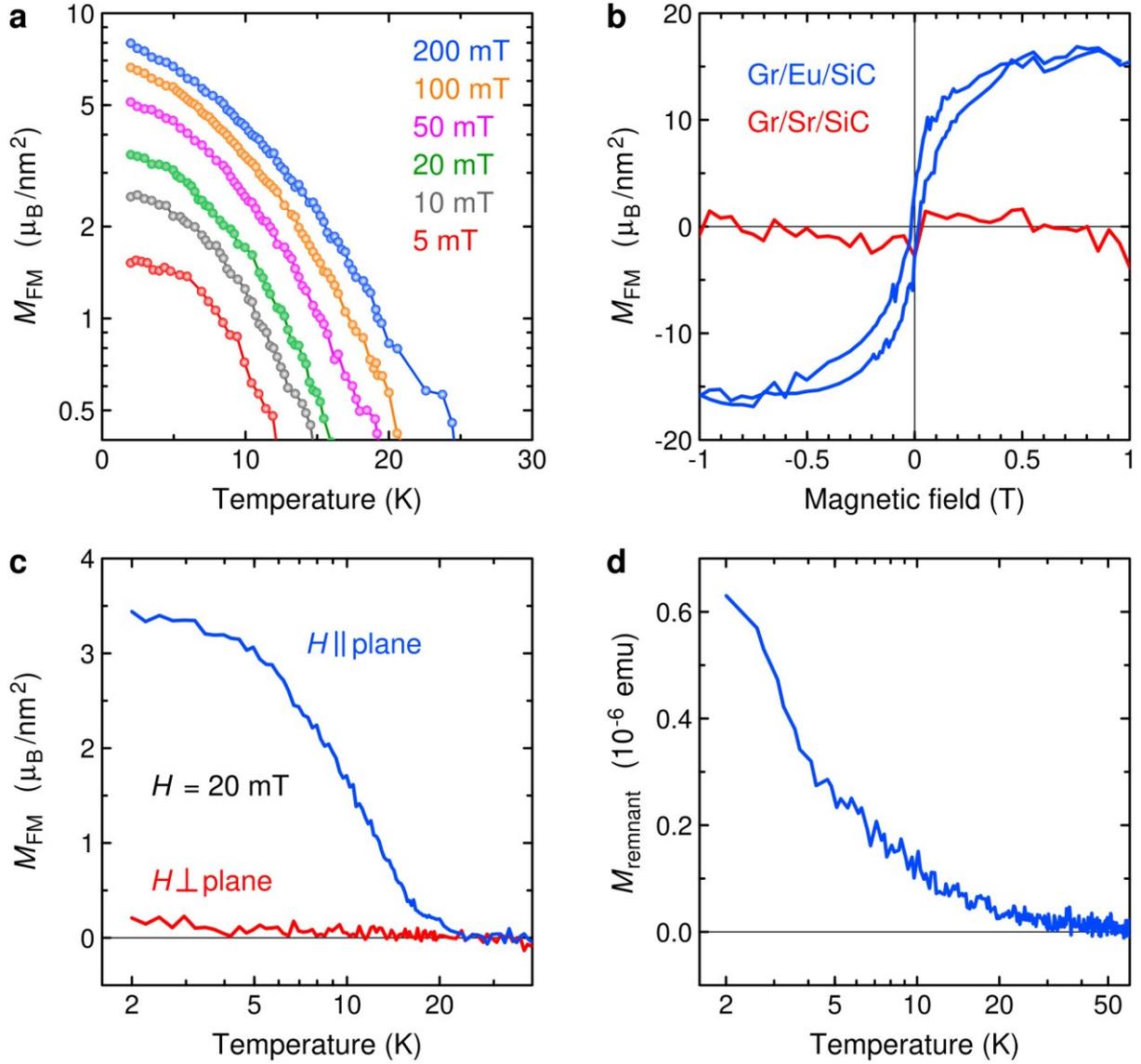

**Fig. 2.** a) Temperature dependence of the area-normalized magnetic moment $M_{FM}$ in Gr/Eu/SiC measured in in-plane magnetic fields 5 mT (red), 10 mT (gray), 20 mT (green), 50 mT (magenta), 100 mT (orange), and 200 mT (blue). b) Dependence of $M_{FM}$ on in-plane magnetic fields at 2 K in Gr/Eu/SiC (blue, *M-H* hysteresis loop shown) and Gr/Sr/SiC (red). c) Magnetic anisotropy in Gr/Eu/SiC: temperature dependence of $M_{FM}$ in a magnetic field 20 mT parallel (blue) or orthogonal (red) to the film surface. d) Temperature dependence of the remnant moment in Gr/Eu/SiC after cooling in an in-plane magnetic field 1 T.

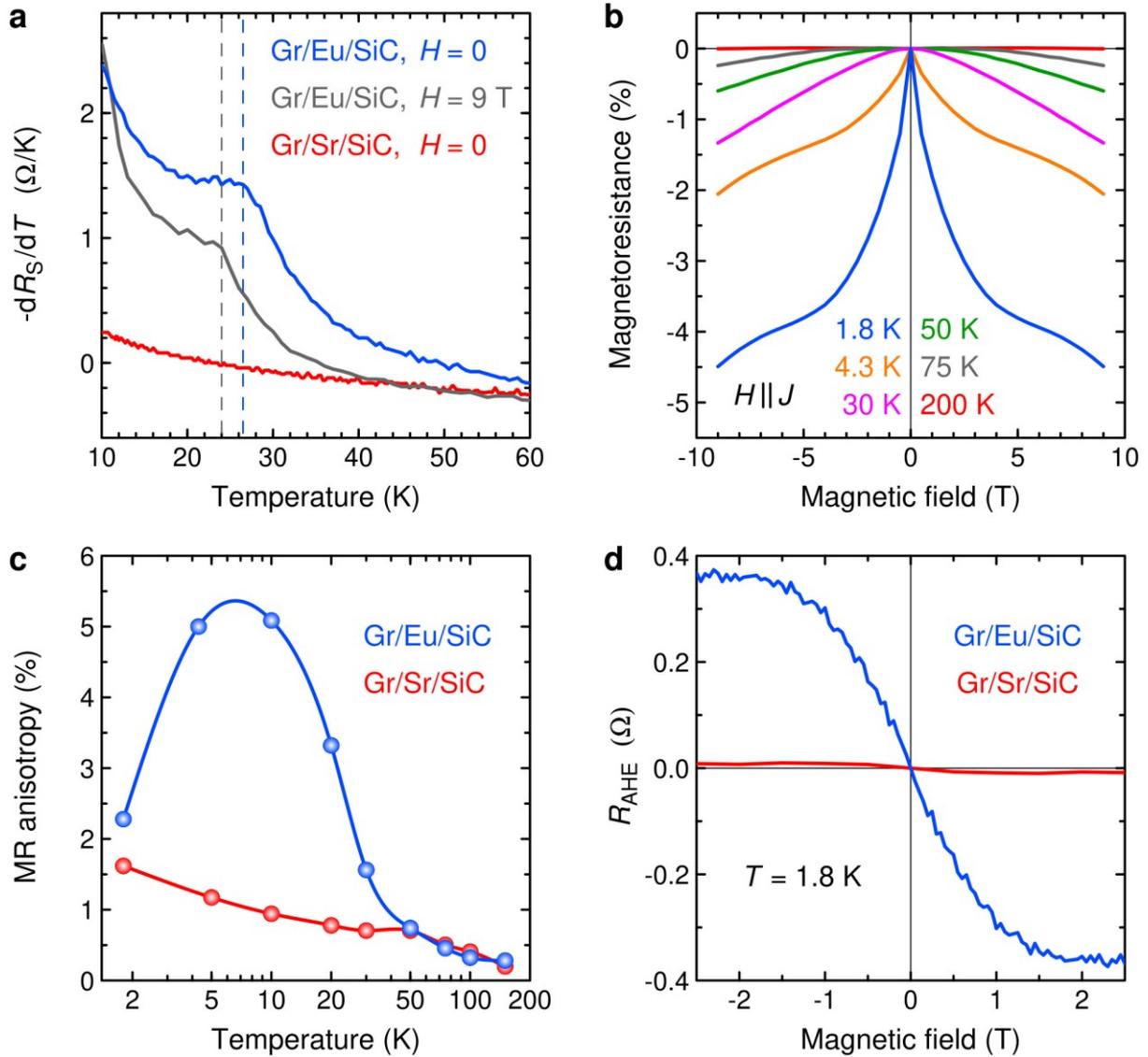

**Fig. 3.** a) Negative shift of the characteristic magnetic transition temperature in a magnetic field: temperature dependence of the derivative of the sheet resistance $R_S$ in Gr/Eu/SiC in zero magnetic field (blue) and in a magnetic field 9 T parallel to the current (gray); for comparison, the featureless dependence in nonmagnetic Gr/Sr/SiC is also shown (red). b) Dependence of MR in Gr/Eu/SiC on a magnetic field parallel to the current at 1.8 K (blue), 4.3 K (orange), 30 K (magenta), 50 K (green), 75 K (gray), and 200 K (red). c) Temperature dependence of the MR anisotropy – the difference between MR in in-plane magnetic fields 9 T parallel and orthogonal to the current – in Gr/Eu/SiC (blue) and Gr/Sr/SiC (red). d) Nonlinear (AHE) contribution to the Hall resistance at 1.8 K in Gr/Eu/SiC (blue) and Gr/Sr/SiC (red).

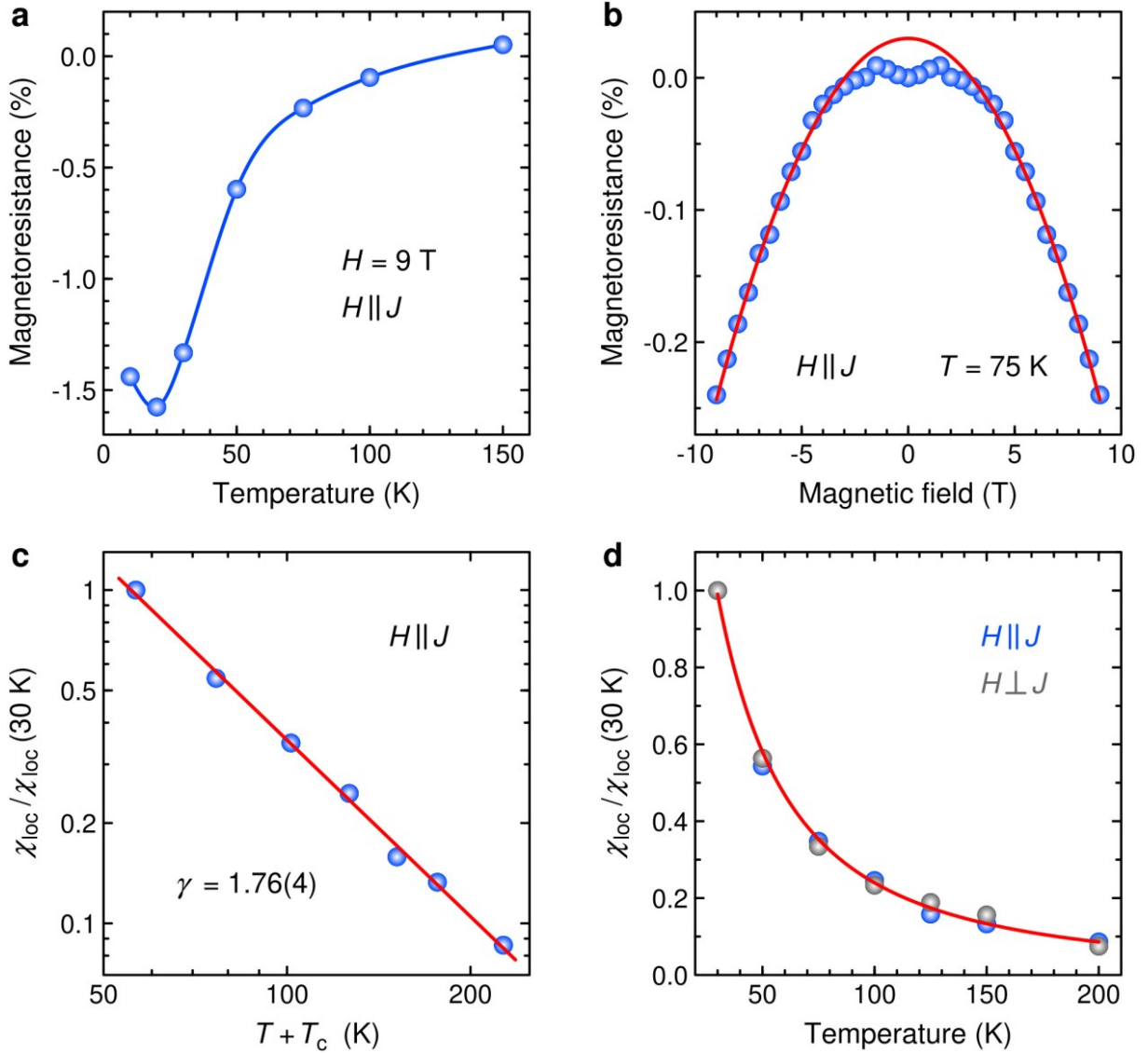

**Fig. 4.** MR in the paramagnetic phase of Gr/Eu/SiC. a) Temperature dependence of MR in a magnetic field 9 T parallel to the current. b) Dependence of MR on magnetic field parallel to the current at 75 K: experimental data (blue dots) and the parabolic fit (red curve). c) Linearized dependence of the normalized local susceptibility $\chi_{loc}$ in magnetic fields parallel to the current on temperature $(T + T_c)$, to determine the critical exponent $\gamma$. d) Comparison of temperature dependences of the normalized local susceptibility $\chi_{loc}$ in in-plane magnetic fields parallel (blue) and orthogonal (gray) to the current; the red curve corresponds to the linear fit in c).

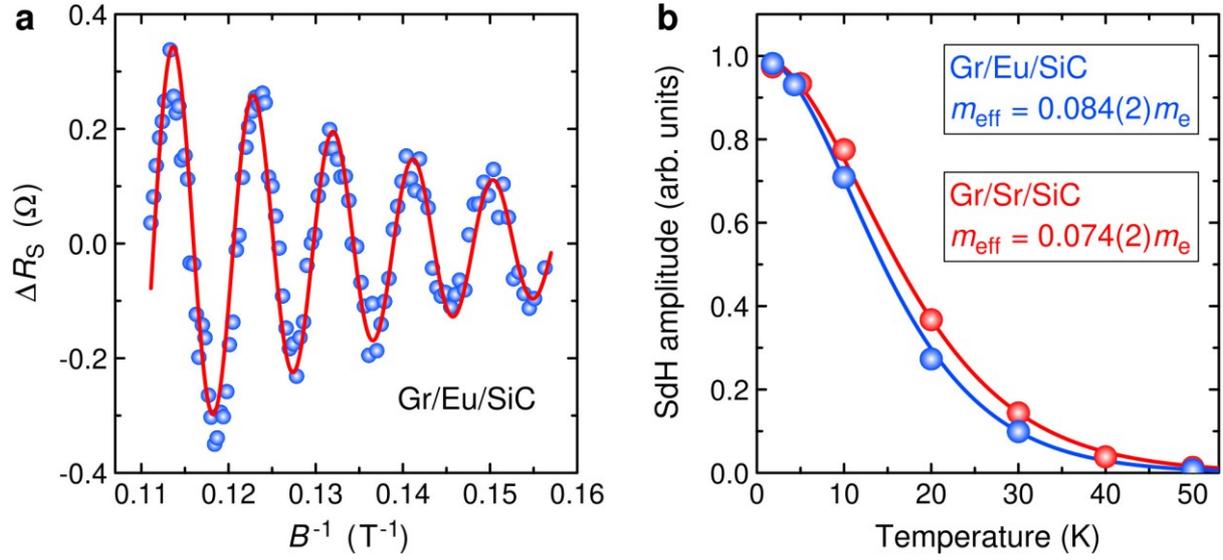

**Fig. 5.** a) Oscillatory part of the sheet resistance of Gr/Eu/SiC at 1.8 K (blue dots). b) Temperature dependence of the normalized amplitude of the SdH oscillations in Gr/Eu/SiC (blue dots) and Gr/Sr/SiC (red dots). Fits in a) and b) correspond to description of the SdH oscillations using the Lifshitz-Kosevich formalism.

# Supporting Information

# Epitaxial Graphene Integrated with a Monolayer Magnet

*Ivan S. Sokolov, Dmitry V. Averyanov, Oleg E. Parfenov, Alexey N. Mihalyuk, Alexander N. Taldenkov, Oleg A. Kondratev, Ilya A. Eliseyev, Sergey P. Lebedev, Alexander A. Lebedev, Andrey M. Tokmachev, and Vyacheslav G. Storchak\**

## Content:

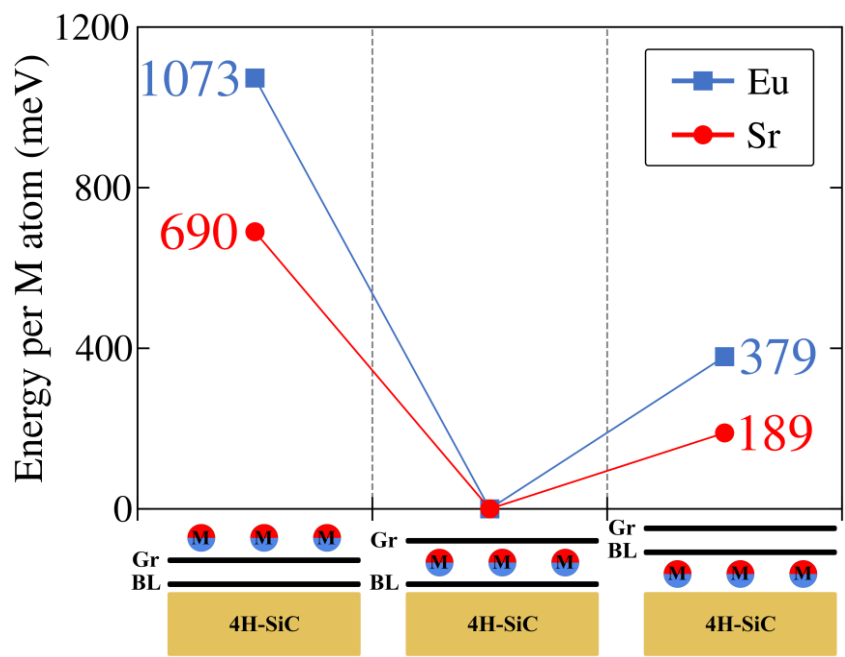

**Fig. S1.** Relative energies per metal atom M (= Eu (blue squares) or Sr (red circles)) of 3 potential $(\sqrt{3} \times \sqrt{3})R30°$ structures of M deposited on 4H-SiC according to DFT calculations: M adsorbed on top of the graphene (Gr) layer, M intercalated between graphene and the buffer layer (BL), and M intercalated underneath the buffer layer. The energy of the most stable structure is set to zero.

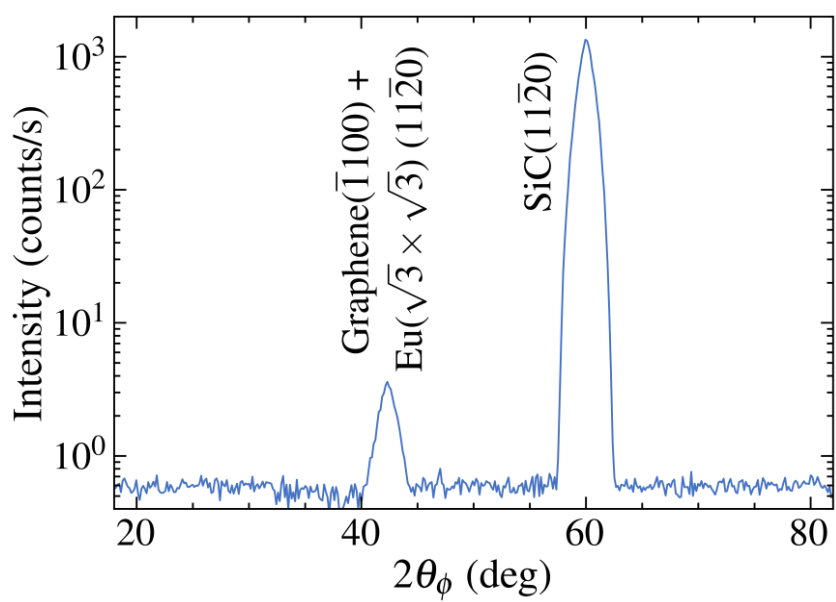

**Fig. S2.** Grazing incidence XRD scan of Gr/Eu/SiC along the direction $[11\bar{2}0]$ of the substrate.

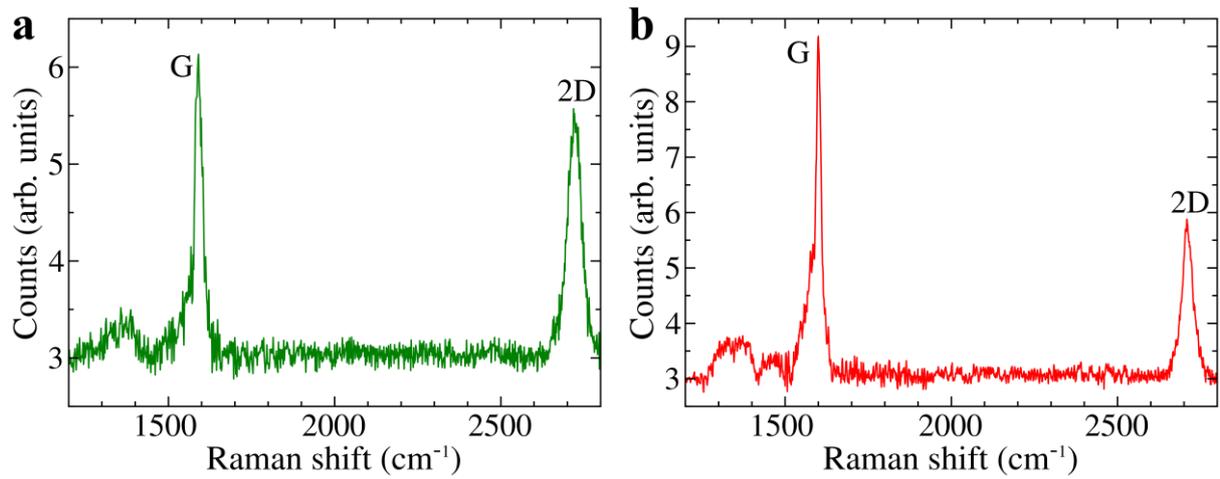

**Fig. S3.** Raman spectra of a) Gr/SiC and b) Gr/Eu/SiC.

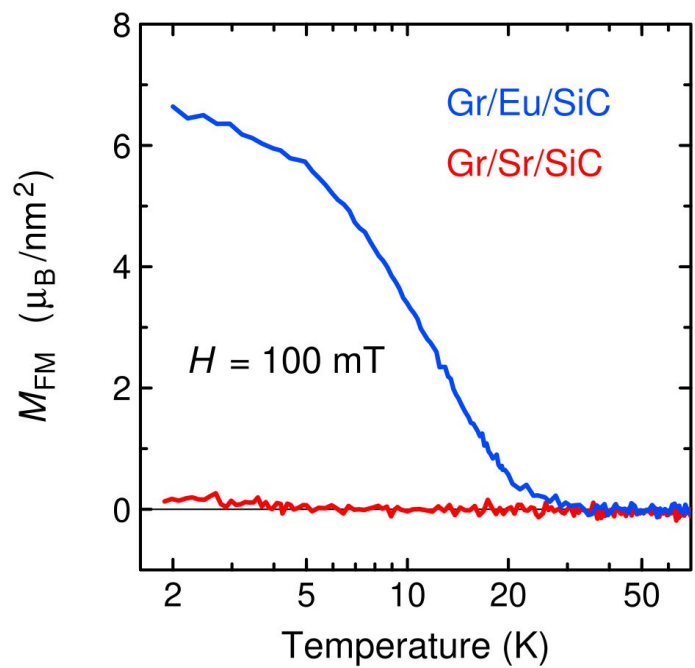

**Fig. S4.** Temperature dependence of the area-normalized magnetic moment $M_{FM}$ in Gr/Eu/SiC (blue) and Gr/Sr/SiC (red) measured in an in-plane magnetic field 100 mT.

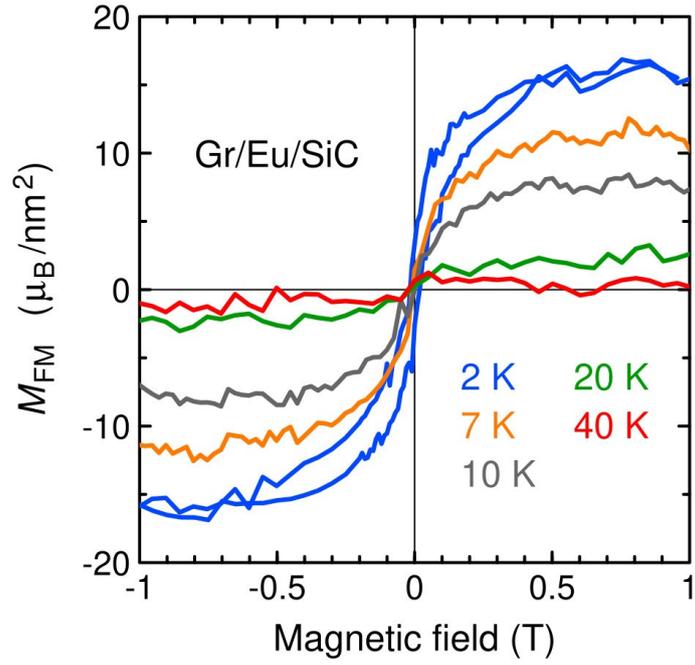

**Fig. S5.** Dependence of $M_{FM}$ in Gr/Eu/SiC on in-plane magnetic fields at 2 K (blue, the *M-H* hysteresis loop shown), 7 K (orange), 10 K (gray), 20 K (green), and 40 K (red).

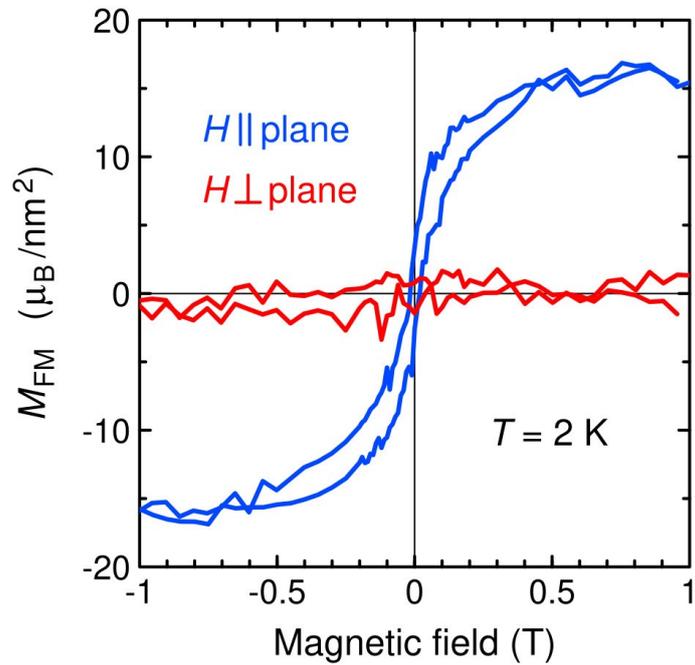

**Fig. S6.** *M-H* curves measured for Gr/Eu/SiC at 2 K in in-plane (blue) and out-of-plane (red) magnetic fields.

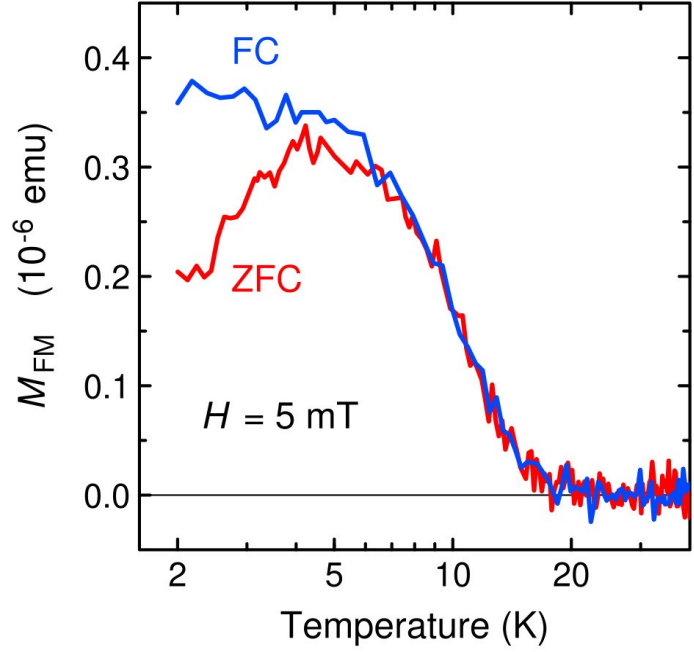

**Fig. S7.** Temperature dependence of the FM magnetic moment in Gr/Eu/SiC for zero-field-cooling (ZFC, red) and field-cooling (FC, blue) in an in-plane magnetic field 5 mT.

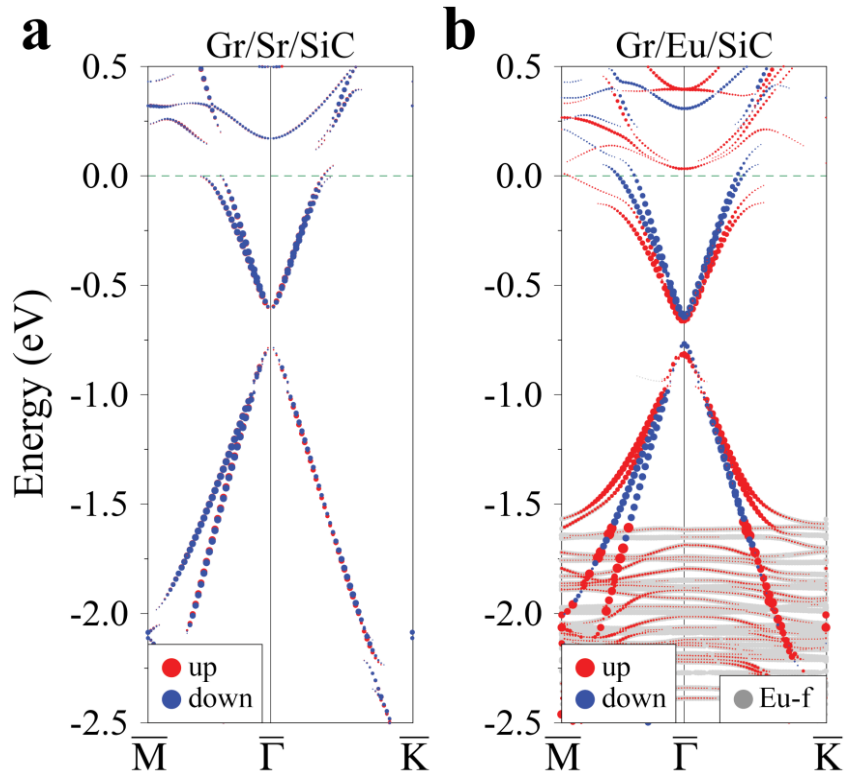

**Fig. S8.** Band structures of a) Gr/Sr/SiC and b) Gr/Eu/SiC; the bands with spins up and down are marked as red and blue, respectively; the $f$-bands of Eu are shown in gray color.

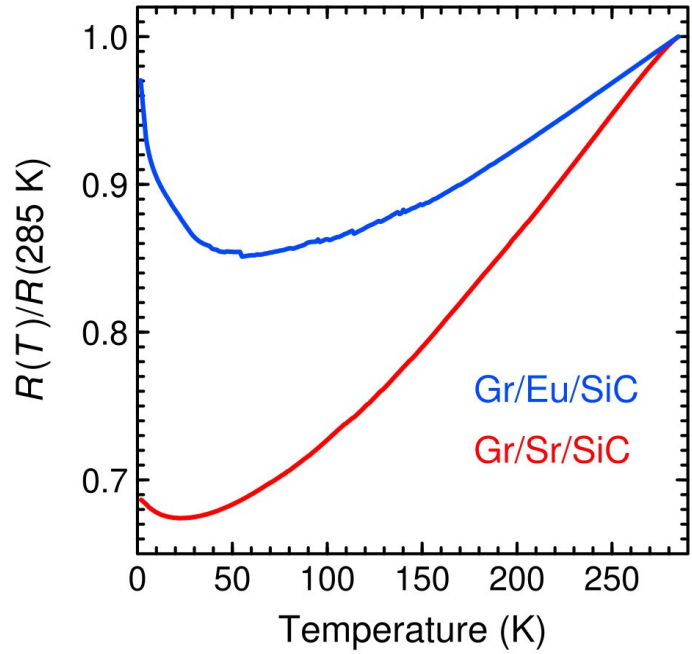

**Fig. S9.** Temperature dependence of the normalized resistance of Gr/Eu/SiC (blue) and Gr/Sr/SiC (red).

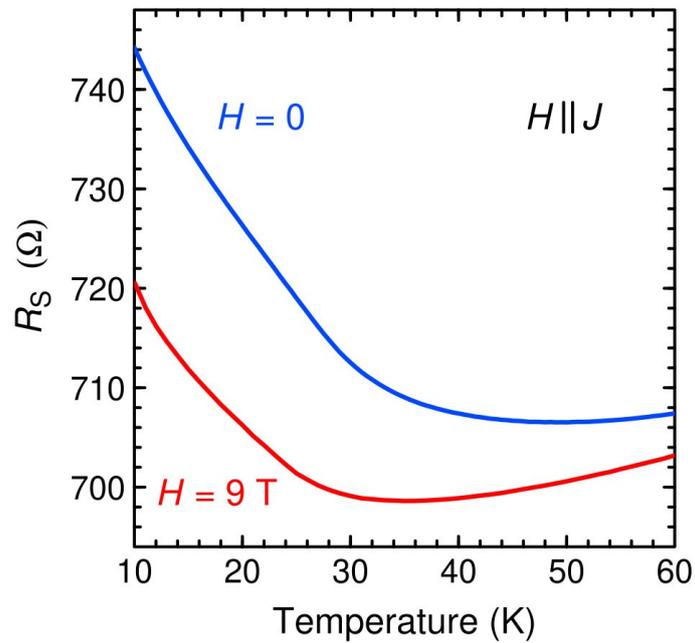

**Fig. S10.** Temperature dependence of the sheet resistance of Gr/Eu/SiC in zero magnetic field (blue) and a magnetic field 9 T parallel to the current (red).

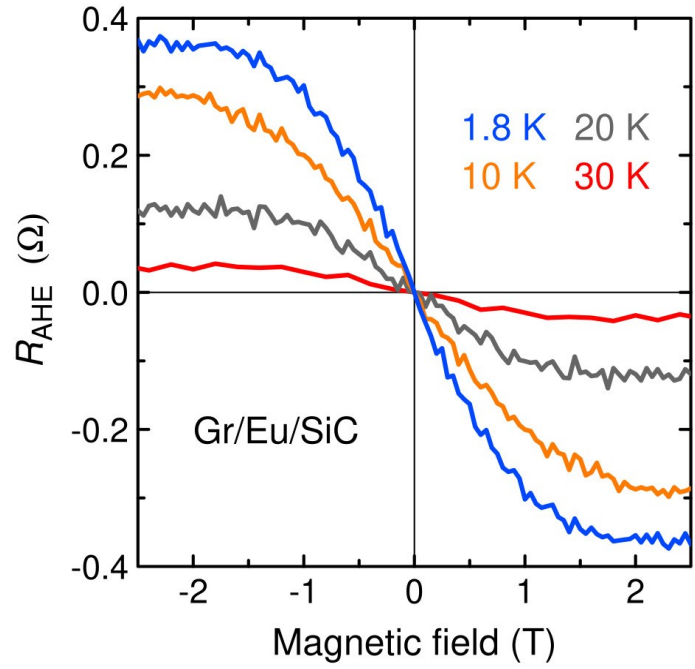

**Fig. S11.** Non-linear (AHE) contribution to the Hall resistance in Gr/Eu/SiC at 1.8 K (blue), 10 K (orange), 20 K (gray), and 30 K (red).

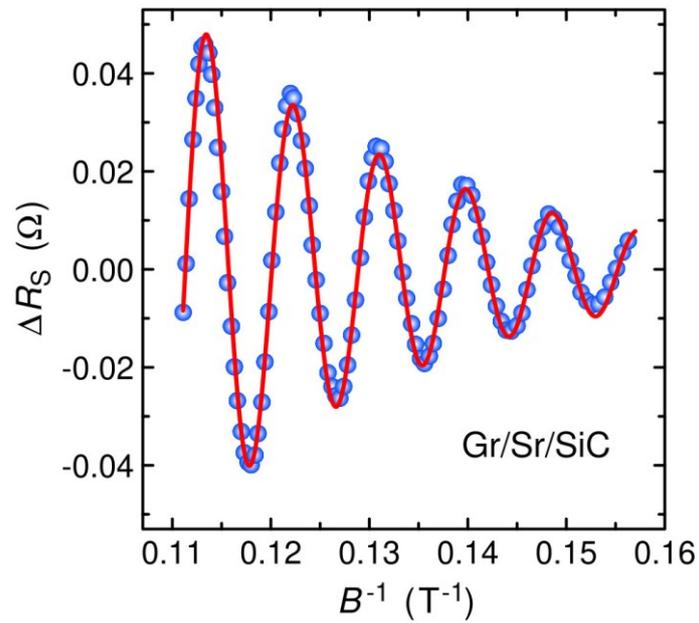

**Fig. S12.** Oscillatory part of the sheet resistance of Gr/Sr/SiC at 1.8 K (blue dots) and its fit by a damped sine wave (red line).